\documentclass[twoside]{llncs}
\usepackage{amsmath}
\usepackage{amssymb}
\usepackage{graphics}
\usepackage[dvips]{epsfig}
\usepackage{times}
\usepackage{fancyhdr}
\usepackage{ifthen}

\newcommand\settitle[2][]{%
 \title{#2}
 \ifthenelse{\equal{#1}{}}%
  {\fancyhead[RO]{\nouppercase #2 \qquad \thepage}}%
  {\fancyhead[RO]{\nouppercase #1 \qquad \thepage}}%
}
\newcommand\setauthors[2]{%
 \author{#2}
  {\fancyhead[LE]{\thepage \qquad \nouppercase #1}}%
}
\fancypagestyle{plain}{%
\fancyhf{}

}

\def\keywordsname{Keywords.}
\newenvironment{keywords}{%
      \list{}{\advance\topsep by-0.50cm\relax\small
     \leftmargin=1cm
      \labelwidth=1cm
     \listparindent=1cm
     \itemindent\listparindent
      \rightmargin\leftmargin}\item[\hskip\labelsep
                                    \bfseries\keywordsname]}

   {\endlist}

   \newtheorem{Remark}{Remark}
   \newtheorem{Rem}{Remark}
   \newtheorem{Def}{Definition}
   \newtheorem{Thm}{Theorem}

   \def\C{{\mathbb C}}
   \def\R{{\mathbb R}}
   
   \def\P{{\mathbb P}}

   \def\Pf{{\it Proof.$\;\;$}}

   \def\spann{\mbox{\rm span}}
   \def\tr{\mbox{\rm tr~}}

   \def\cM{{\mathcal M}}
   \def\cQ{{\mathcal Q}}
   
   \def\cV{{\mathcal V}}
   \def\cP{{\mathcal P}}

\def\C{{\mathbb C}}

\def\P{{\mathbb P}}
\def\R{{\mathbb R}}

\def\Pf{{\it Proof.$\;$}}

\def\qed{\hspace{10cm}$\diamond$}
\def\qedd{\hfill$\diamond$}

\def\diag{\mbox{\rm diag}}

\def\Pr{\mbox{\rm Pr}}

\def\tr{\mbox{\rm tr}}

\def\({\langle}
\def\){\rangle}
\def\mb{\boldsymbol}
\def\im{{\rm i}}

\def\cC{{\mathcal C}}
\def\cD{{\mathcal D}}

\def\cH{{\mathcal H}}

\def\cM{{\mathcal M}}
\def\cN{{\mathcal N}}

\def\cP{{\mathcal P}}
\def\cQ{{\mathcal Q}}
\def\cR{{\mathcal R}}

\def\cV{{\mathcal V}}
\def\cW{{\mathcal W}}

\def\1{\mb1}

\def\v0{{\bf 0}}

\def\ve{{\bf e}}

\def\vq{{\bf q}}

\def\vP{{\bf P}}

\def\ov{\overline}

\begin{document}

\settitle[]{On Hidden States in Quantum Random Walks}

\setauthors{U. Faigle/A. Sch\"onhuth}
           {Ulrich Faigle$^1$ and Alexander Sch\"onhuth$^2$}
\institute{Mathematisches Institut\\
           Universt\"at zu K\"oln\\
           Weyertal 80\\ 50931 K\"oln, Germany\\
\email{faigle@zpr.uni-koeln.de}\\[1ex]
\and
           Centrum Wiskunde \& Informatica\\
           Science Park 123\\
           1098 XG Amsterdam, The Netherlands\\
\email{A.Schoenhuth@cwi.nl}}

\date{}
\maketitle

\thispagestyle{plain}
\begin{abstract}
  It was recently pointed out that identifiability of quantum random
  walks and hidden Markov processes underlie the same principles.
  This analogy immediately raises questions on the existence of hidden
  states also in quantum random walks and their relationship with
  earlier debates on hidden states in quantum mechanics.  The
  overarching insight was that not only hidden Markov processes, but
  also quantum random walks are finitary processes. Since finitary
  processes enjoy nice asymptotic properties, this also encourages to
  further investigate the asymptotic properties of quantum random
  walks.  Here, answers to all these questions are given. Quantum
  random walks, hidden Markov processes and finitary processes are put
  into a unifying model context. In this context, quantum random walks
  are seen to not only enjoy nice ergodic properties in general, but
  also intuitive quantum-style asymptotic properties. It is also
  pointed out how hidden states arising from our framework relate to
  hidden states in earlier, prominent treatments on topics such as the
  EPR paradoxon or Bell's inequalities.
\end{abstract}

\section{Introduction \label{intro}}

Quantum random walks were introduced in 2001~\cite{Aharonov01}, as a
concept that can emulate Markov chain based techniques on quantum
computers~\cite{Kempe03}. This analogy in terms of application
immediately raises questions relating to theoretical analogies. Do
quantum random walks have nice asymptotic properties, such as
favorable convergence rates?  And, when relating quantum random walks
to Markovian latent variable models such as hidden Markov processes:
are there any reasonable latent variables also in quantum random
walks? And, if so, can one perform convenient computations on
those hidden states?

Recent research~\cite{Faigle11} pointed out that both hidden Markov
processes and quantum random walks are \emph{finitary}. Finitary
processes have been key to determining the equivalence of two
differently parametrized hidden Markov processes and, as became clear
in \cite{Faigle11}, also of two quantum random walks. As was pointed
out in \cite{Schoenhuth09}, finitary processes are also key to
determining ergodicity of those processes, in polynomial time with
respect to the input parameters. These issues are closely
related with the identifiability problem one commonly encounters in
latent variable
modeling~\cite{Blackwell57,Dharma65,Heller65,Ito92}. So these findings
put quantum random walks immediately into the focus of questions
concerning hidden variables.

Finitary processes can be viewed as acting on underlying hidden states
that have been decoupled from probability theory---while hidden states
still exist, there is no probabilistic prescription for how they
operate~\cite{Heller65,Faigle07}. While this renders it impossible to
estimate in which of those hidden states the system is actually in,
this comes with non-negligible benefits in compensation. As
above-mentioned, one creates a frame that allows for determining
equivalence and ergodicity. Moreover, it allows for (sometimes
dramatic) reductions in terms of model complexity. There are examples
of stochastic processes, quite intuitively based on only few
underlying hidden states, that require an infinite number of hidden
states when stipulating probabilistic interpretation in addition (the
``probability clock''; see \cite{Jaeger00}). However, they indeed are
based on just the intuitive, finite number of hidden states, if one
gets rid of these probabilistic constraints.

The purpose of this paper is to thoroughly explore these
relationships.  We provide a formal frame that puts hidden Markov
processes, quantum random walks and finitary processes into one,
unifying context. This frame allows us to prove convenient,
quantum-style ergodic properties (\emph{``stationary limit
  densities''}) for quantum random walks first of all. As a sound
justification of our doing, our framework allows us to point out a
natural (and, as we feel, quite striking) analogy between finitary
processes on the one hand, and the corresponding counterpart emerging
from quantum random walks on the other hand. In short, we demonstrate
that freeing hidden states from probability theory in the context of
stochastic process theory is equivalent to freeing hidden states from
being measurable in the context of quantum mechanical counterparts of
finitary processes, as a generalization of quantum random walks.

We finally point out that our framework can also draw a connection to
(historically prominent) debates on the existence of hidden states
within the frame of the quantum mechanics formalism. Examples and
results raised around those debates---the EPR paradox and Bell's
inequalities, for example
\cite{Bell64,Bell66,Einstein,Mueckenheim,deMuynck,Scully-Walther-Schleich}---can
be conveniently rephrased using our framework, which allows to
maintain a clear, formal view on possible hidden states in this
context.

\section{Preliminaries}
\label{sec:Preliminaries}

\subsection{Hermitian Matrices}
$\R$ denotes the scalar field of real numbers and $\C$ the field of
complex numbers $z= a+\im b$ (where $a,b\in \R$ and $\im^2
=-1$). $\C^{m\times n}$ is the ($mn$)-dimensional vector space of all
$(m\times n)$-matrices of the form
\begin{equation}\label{eq.matrix-form}
     C = A +\im B \quad\mbox{with $A,B\in \R^{m\times n}$.}
\end{equation}
$\ov{C} = A -\im B$ is the \emph{conjugate} of $C=A+\im B$. The
transpose $C^* = \ov{C}'$ of its conjugate $\ov{C}$ is the
\emph{adjoint} of $C$. $\C^{m\times n}$ is a Hilbert space with
respect to the \emph{Hermitian inner product}
\begin{equation}
\(C|D\) :=  \tr(C^*D) =\sum_{i=1}^m\sum_{j=1}^n \ov{c_{ji}}d_{ij},
\end{equation}
where the $c_{ij}$ and the $d_{ij}$ denote the coefficients of $C$ and
$D$. $\|C\|:= \sqrt{\(C|C\)}$ is the \emph{norm} of $C$.  $\C^n$ is
short for $\C^{n\times 1}$ and can also be identified with the space
of diagonal matrices in $\C^{n\times n}$:
\begin{equation}
\begin{bmatrix}
u_1\\
u_2\\
\vdots \\
u_n
\end{bmatrix} \in \C^n \quad\longleftrightarrow\quad \diag(u_1,u_2,\ldots,u_n) =
\begin{bmatrix}
u_1 &0 &0 &\ldots\\
0 &u_2 &0 &\ldots\\
\vdots& &\ddots\\
0& &\ldots &u_n
\end{bmatrix}
\end{equation}

\medskip
Assuming $m=n$, a matrix $C=A+\im B$ with the property $C^*=C$ is
\emph{self-adjoint} or \emph{Hermitian}, which means that $A$ is
\emph{symmetric} (\emph{i.e.}, $A^T=A$) and $B$ \emph{skew-symmetric}
(\emph{i.e.}, $B^T = -B$).  Let $\cH_n$ denote the collection of all
Hermitian $(n\times n)$-matrices. From the general form
(\ref{eq.matrix-form}) one recognizes $\cH_n$ as a real Hilbert space
of dimension $\dim_\R(\cH_n) = n^2$.

\medskip
A matrix $Q=[q_{ij}]\in \cH_n$ has real eigenvalues $\lambda_i$ and a
corresponding orthonormal set $\{u_1,\ldots,u_n\}$ of eigenvectors
$u_i\in \C^n$, yielding the \emph{spectral decomposition}
\begin{equation}\label{eq.Spektraldarstellung}
Q = \sum_{i=1}^n \lambda_i {u}_i u^*_i \quad\mbox{and hence trace}\quad
   \tr(Q) = \sum_{i=1}^n q_{ii} = \sum_{i=1}^n \lambda_i.
\end{equation}
$Q\in \cH_n$ is said to be \emph{nonnegative} if all eigenvalues of
$Q$ are nonnegative, which is equivalent to the property
\begin{equation}\label{eq.nonnegativity}
    u^*Qu \geq 0 \quad \mbox{holds for all $u\in \C^n$.}
\end{equation}
A vector $u\in \C^n$ gives rise to a nonnegative element
$uu^*\in\cH_n$ and one has
\begin{equation}\label{eq.length-trace}
  \(u|u\) = \tr(uu^*).
\end{equation}
In the case $\tr(uu^*) = 1$, the matrix $uu^*$ is thought to represent
a \emph{pure state} of an $n$-dimensional quantum system.

\subsection{Unitary Operators}
A matrix $U\in \C^{n\times n\\}$ is \emph{unitary} if the identity
matrix $I$ factors into $I=UU^*$, \emph{i.e.}, if the row (or column)
vectors of $U$ form an orthonormal basis for $\C^n$. So also $U^*$ is
unitary. For example, an orthonormal basis $\{u_1,\ldots, u_n\}$ of
eigenvectors relative to $Q\in \cH_n$ gives rise to a unitary matrix
$U^*$ with columns $u_i$. Where $\lambda_1,\ldots,\lambda_n$ are the
corresponding eigenvalues, the linear operator $x\mapsto Qx$ on $\C^n$
is described with respect to the basis $U^*$ {\it via} the transformed
matrix
\begin{equation}\label{eq.Basiswechsel}
    UQU^* = \diag(\lambda_1,\ldots,\lambda_n) \in \cH_n.
\end{equation}

\subsection{Strings and Process Functions.}
\label{sec.sp}
Let $\Sigma$ be a finite alphabet. We write $a,b\in\Sigma$
for single letters and $v,w\in\Sigma^* = \cup_{t\ge 0}\Sigma^t$ for
strings, where $\Sigma^0=\{\epsilon\}$ with $\epsilon$ the
\emph{empty string}. Concatenation of 
$v=v_1...v_t\in\Sigma^t,w=w_1...w_s\in\Sigma^s$ is written
$vw=v_1...v_tw_1...w_s\in\Sigma^{t+s}$. We consider stochastic
processes $(X_t)$ taking values in $\Sigma$ as string functions
$p:\Sigma^*\to\R$ where
\begin{enumerate}
\item\label{eq.sp1} $p(v)\ge 0$ for all $v\in\Sigma^*$,
\item\label{eq.sp2} $\sum_{a\in\Sigma}p(va)=p(v)$ for all $v\in\Sigma^*$,
\item\label{eq.sp3}  $p(\epsilon)=1$.
\end{enumerate}
Such string functions
are in one-to-one correspondence with stochastic processes via the
relationship [for technical convenience, stochastic processes
  start at $t=1$]
\begin{equation}
  \vP(\{X_1=v_1,...,X_t=v_t\}) = p(v_1...v_t)
\end{equation}
due to standard measure-theoretic arguments.
We refer to such string functions $p$ as \emph{process
  functions}.

\section{Processes}
\label{sec.processes}

In the following, we will identify \emph{quantum random walks (QRWs)}
with the stochastic processes associated with it and we will refer to
their parametrizations as QRW parametrizations. Furthermore, we will
distinguish between \emph{hidden Markov processes (HMPs)} and
\emph{hidden Markov models (HMMs)}, where the latter are the
parametrizations of HMPs.

\medskip
We summarize these facts for introductory purposes only; none of what
follows in this section is new. See the citations listed in the
following for more details.

\subsection{Finitary Processes}
\label{sec.finitary}

As was pointed out in \cite{Faigle11}, both \emph{hidden Markov
  processes (HMPs)} and \emph{quantum random walks (QRWs)} are
\emph{finitary processes}.  While this was new for QRWs, this was well
known for HMPs. Finitary processes emerged in early work on HMP
identification (e.g.~\cite{Blackwell57,Dharma65,Gilbert59,Heller65})
and have remained a core concept also in recent work on
identifiability~\cite{Finesso10,Ito92,Schoenhuth14,Vidyasagar09}.
Finitary processes are sometimes also referred to as {\em linearly
  dependent} \cite{Ito92}, {\em observable operator models}
\cite{Jaeger00} or as {\em finite-dimensional}
\cite{Faigle07,Schoenhuth09}. In their possibly most prevalent
application they served to determine equivalence of hidden Markov
processes (HMPs) in 1992 \cite{Ito92}. The exponential runtime
algorithm was later improved to polynomial runtime~\cite{Faigle11}.

\begin{Def}[Finitary Process]
\label{d.finitary}
A stochastic process $p:\Sigma^*\to\R$ is said to be \emph{finitary} iff there
are matrices $M_a\in\R^{d\times d}$ for all $a\in\Sigma$ and a vector
$\pi\in\R^d$ where [let $^T$ denote matrix transposition and
  $\mathbf{1}$ be the vector of all ones]
\begin{enumerate}
\item $M:=\sum_aM_a$ has unit row sums, i.e.~$M\mathbf{1}=\mathbf{1}$ and 
\item $\pi$ is a unit vector, i.e.~$\pi^T\mathbf{1}=1$
\end{enumerate}
such that 
\begin{equation}
\label{eq.finitary}
p(v_1...v_t) = \pi^TM_{v_1}\cdot\ldots\cdot M_{v_n}\mathbf{1}
\end{equation}
Because of $\pi\in\R^d$ and $M_a\in\R^{d\times d}$ for all
$a\in\Sigma$, the parametrization $((M_a)_{a\in\Sigma},\pi)$ is
referred to as \emph{$d$-dimensional}.
\end{Def}

It is an immediate observation that a finitary process that admits a
$d$-dimensional parametrization also admits a parametrization of
dimension $d+1$, which allows for the following definition.

\begin{Def}[Rank of a Finitary Process]
\label{d.finitaryrank}
The \emph{rank} of a finitary process $(X_t)$ is the minimal dimension
of a parametrization that it admits.
\end{Def}

\subsection{Hidden Markov Processes}
\label{sec:hmms}

A \emph{hidden Markov process (HMP)} is parametrized by a tuple $\cM=(S, E,
\pi, M)$ where
\begin{enumerate}
\item $S=\{s_1,\ldots,s_n\}$ is a finite set of \emph{``hidden'' states}
\item $E=[e_{ia}]\in\R^{S\times \Sigma}$ is a non-negative {\em emission
  probability matrix} with unit row sums $\sum_{a\in\Sigma}e_{ia}=1$,
  (\emph{i.e.} the row vectors of $E$ are probability distributions on
  $\Sigma$)
\item $\pi$ is an \emph{initial probability distribution} on $S$ and 
\item $M=[m_{ij}]\in\R^{S\times S}$ is a non-negative \emph{transition
  probability matrix} with unit row sums $\sum_{i=1}^nm_{ij} =1$
  (\emph{i.e.} the row vectors of $M$ are probability distributions on
  $S$)
\end{enumerate}

The associated process $(X_t)$ initially moves to a state $s_i\in S$
with probability $\pi_i:=\pi_{s_i}$ and emits the symbol $X_1=a$ with
probability $e_{ia}$. Then it moves from $s_i$ to a state $s_j$ with
probability $m_{ij}$ and emits the symbol $X_2=a'$ with probability
$e_{ja'}$ and so on. In the following, we also refer to a
parametrization $\cM=(S,E,\pi,M)$ as a \emph{hidden Markov model
  (HMM)}. See \cite{Ephraim02} for a comprehensive review.

\begin{Remark}
  \label{rem.ffmc}
  Replacing the emission probability matrix $E$ by a function
  $f:S\to\Sigma$, which models that from hidden state $s$ the value
  $f(s)$ is observed with probability one, gives rise to a class of
  processes referred to as \emph{finite functions of Markov chains (FFMCs)}.
  It is relatively straightforward to observe (see \cite{Ito92}) that
  the class of hidden Markov processes is equivalent to that of FFMCs.
\end{Remark}
\subsubsection{HMPs are finitary}

HMPs $p:\Sigma^*\to\R$ are immediately shown to be finitary by the
observation that the \emph{transition matrix} $M\in\R^{S\times S}$
decomposes as 
\begin{equation}
  \label{eq.hmpdecomp}
  M=\sum_{a\in \Sigma} M_a,\quad\text{with
coefficients}\quad
    (M_a)_{ij} := e_{ia}\cdot m_{ij}.
\end{equation}
These coefficients reflect the probabilities to emit symbol $a$ from
state $s_i$ and to move on to state $s_j$. Standard
technical computations then indeed yield that
\begin{equation}
  p(v_1...v_t) = \pi^TM_{v_1}\ldots M_{v_{t-1}}M_{v_t}\mathbf{1}.
\end{equation}
This shows $p$ to be a finitary process of rank at most $|S|$,
the number of hidden states.

\begin{Rem}
  \label{rem.finitary-vs-hmp}
  HMPs on d hidden states of rank d and finitary processes that do not
  admit a HMM parametrization are known to exist. See, for example,
  Ex.~3.8 in \cite{Schoenhuth14} for the former and see
  \cite{Jaeger00} for the latter, where the ``probability clock'' has
  rank 3 as finitary process, but only admits a HMP formulation on
  an infinite number of hidden states.
\end{Rem}

\subsection{Quantum Random Walks}
\label{sec:qrws}

In earlier work of ours \cite{Faigle11}, we had pointed out a
connection between \emph{quantum random walks (QRWs)} and finitary
processes. We will briefly revisit this connection here for the sake
of illustration. In the following, we consider a QRW as given by a
unitary operator $U$ together with an initial wave function
$\psi_0$. Usually, as per a most general definition, the QRW
$(U,\psi_0)$ is supposed to reflect the locality structure of $P$, the
probability matrix of a discrete-time Markov chain \cite{Szegedy04}.

\paragraph{\bf Szegedy's Model.}
Note that the following example connects finitary processes with a
quantum walk model that was raised in a seminal paper
\cite{Aharonov01}.  In the meantime, several reformulations of QRWs
have been raised, including the popular and attractive one by Szegedy
\cite{Szegedy04}.  We note already here that also Szegedy's model is
covered by our treatment.  However, while this connection is even
easier to draw than for Aharonov et al.'s model \cite{Aharonov01}, it
requires to raise the definition of \emph{Quantum Markov Chains}
first. See subsection \ref{ssec.qmcs-and-qrws} for the corresponding
arguments.

\medskip
\paragraph{\bf Aharonov's Early Model.}
In the seminal work of \cite{Aharonov01} (see also \cite{Kempe03}), a
\emph{quantum random walk (QRW)} is parametrized by a tuple
$\cQ=(G,U,\psi_0)$ where
\begin{enumerate}
\item $G=(\Sigma,E)$ is a directed graph over the alphabet $\Sigma$
\item $U:\C^k\to\C^k$ is a unitary \emph{evolution} operator where
  $k:=|E|=K\cdot|\Sigma|$ and
\item $\psi_0\in\C^k$ is a wave function, i.e.~$||\psi_0||=1$
  [$||.||$ is the Euclidean norm].
\end{enumerate}
Edges are labeled by tuples $(a,x), a\in\Sigma, x\in X$ where $X$ is
a finite set with $|X|=K$. Correspondingly, $\C^k$ is considered to be
spanned by the orthonormal basis 
\begin{equation*}
\langle\; \ve_{(a,x)}\,|\,(a,x)\in E \;\rangle. 
\end{equation*}
Following the definition of a \emph{general quantum walk} suggested by
\cite{Aharonov01}, the unitary operator $U$ is supposed to respect the
structure of the graph. That is, if $\cN(a):=\{a'\in\Sigma\mid \exists
(a,a')\in E\}$ are the neighboring nodes of $a$, so
\begin{equation*}
  U(\ve_{(a,x)})\subset\spann\{\ve_{(a',x)}\mid a'\in\cN(a)\cup\{a\}\}.
\end{equation*}

The \emph{quantum random walk} $(X_t)$ arising from a parametrization
$\cQ=(G,U,\psi_0)$ proceeds by first applying the unitary operator $U$
to $\psi_0$ and subsequently, with probability $\sum_{x\in
  X}|(U\psi_0)_{(a,x)}|^2$, ``collapsing'' (i.e.~projecting and
renormalizing, which models a quantum mechanical measurement)
$U\psi_0$ to the subspace spanned by the vectors $\ve_{(a,x),x\in X}$
to generate the first symbol $X_1=a$. Collapsing $U\psi_0$ results in
a new wave function $\psi_1$. Applying $U$ to $\psi_1$ and collapsing
it, with probability $\sum_{x\in X}|(U\psi_1)_{(a',x)}|^2$, to the
subspace spanned by $\ve_{(a',x),x\in X}$ generates the next symbol
$X_2=a'$. Iterative application of $U$ and subsequent collapsing
generates further symbols.

\subsubsection{Quantum random walks are finitary.}
\label{sec:qrws-finitary}

Exposing QRWs as finitary, as per the arguments raised in
\cite{Faigle11}, is based on a fundamental theorem for finitary
processes.

\begin{Def}[Hankel matrix]
Let
\begin{equation}
  \label{eq.hankel}
  \cP_p := [p(vw)_{v,w\in \Sigma^*}]\in\C^{\Sigma^*\times\Sigma^*}
\end{equation}
be the {\em Hankel matrix} of a process function
$p:\Sigma^*\to\R$.
\end{Def}

\noindent Note that both rows and columns of $\cP_p$
\begin{equation*}
  p_v: w\mapsto p(vw)\quad\text{and}\quad p^w:v\mapsto p(vw)
\end{equation*}
are string functions in their own right. Note further that
$\frac{1}{p(v)}p_v$ is a process function for $p(v)>0$, while this is
not necessarily the case for columns $p^w$. We refer to row and column
space of $\cP_p$ as
\begin{equation*}
  \cR(p) = \spann\{p_v\mid v\in\Sigma^*\}\quad\text{and}\quad\cC(p) = \spann\{p^w\mid w\in\Sigma^*\}
\end{equation*}
respectively. Note that, in comparison to earlier work
(\cite{Faigle11}), we have exchanged rows and columns, for the sake of
a more convenient notation.

\medskip
One can show that finitary processes are precisely the processes whose
Hankel matrices have finite rank. In fact, the rank of $\cP_p$ is just
the rank of $p$ as a finitary process.

\begin{Thm}[\cite{Jaeger00,Schoenhuth14}]
\label{t.oom}
Let $p:\Sigma^*\to\C$ be a process function.
Then the following conditions are equivalent.
\begin{enumerate}
\item[(i)] $\cP_p$ has rank at most $d$.
\item[(ii)] There exists a vector $\pi\in\C^d$ and matrices
  $M_a\in\R^{d\times d}$ for all $a\in\Sigma$ such that
  \begin{equation}
    p(a_1...a_n) = \pi^TM_{a_1}\cdot\ldots\cdot M_{a_n}\mathbf{1}
  \end{equation}
  for all $v=a_1...a_n\in\Sigma^*$.
\end{enumerate}
\end{Thm}

The arguments put forward in \cite{Faigle11} proceeded further by
showing that QRWs $p:\Sigma^*\to\R$ allow for choosing a finite number
of string functions $q_1,...,q_{k^2}$, where $k$ is the number of
edges of the graph that underlies the QRW $p$, such that
\begin{equation}
  \cR(p) = \spann\{q_i\mid i=1,...,k^2\}.
\end{equation}
That is, the $q_i$ span the row space of $\cP_p$, the Hankel matrix
of the QRW $p$, which exposes $p$ as a finitary process of rank at
most $k^2$.

\section{Quantum Markov Chains}
\label{sec:qmcs}

We have just seen (section~\ref{sec:qrws-finitary}) that QRWs are
finitary \cite{Faigle11}. As a consequence, QRWs come with some
convenient properties that have been raised for this class of
processes \cite{Faigle07,Faigle11,Schoenhuth09,Schoenhuth09a}.

Convergence rates are a critical issue for QRWs, because QRWs are
supposed to emulate Markov
Chain Monte Carlo (MCMC) based techniques on quantum computers.
This explains why in \cite{Aharonov01} it was pointed out
that the limits
\begin{equation}
  \label{eq.ahalim}
  \bar{p}(v):=\lim_{t\to\infty}\sum_{\bar{w}\in\Sigma^t}p(\bar{w}v)
\end{equation}
for $p$ a QRW (as process function), and $v$ a single letter (node),
exist. This justifies QRWs as a reasonable quantum computational
concept.

\medskip
Finitary processes were shown to be \emph{asymptotically mean
  stationary (AMS)}, see \cite{Faigle07}. Therefore, they have good
ergodic properties (see \cite{Gray,Gray1}). An immediate consequence
of this is, for example, that one can replace single letters $v$ by
arbitrary cylinder sets of strings in \eqref{eq.ahalim}.  Also, the
conditional entropies of AMS processes were shown to converge to a
limit
\begin{equation}
  H_\infty(X)= \lim_{t\to \infty}\frac1t H(X^t) = \lim_{t\to\infty} H(X^t|X^{t-1})
\end{equation}
see \cite{Gray}.  In summary, this already significantly generalizes
\eqref{eq.ahalim}.

\medskip
However, we have not shown limits to exist that have meaning in terms
of the QM formalism, and not only in terms of the statistics derived
from QRWs. As an illustration for the inherent difficulties, let
$(U,\psi_0)$ be a QRW where $U$ does not have $1$ as eigenvalue,
which implies
\begin{equation}
  \label{eq.zerolimit}
  \lim_{t\to\infty}\frac1t\sum_{k=0}^{t-1}U^k\psi_0 = 0.
\end{equation}
Our wishful thinking, however, was to obtain non-trivial \emph{wave
  functions} $\bar{\psi}$ as limits.  While this is not possible, we
will be able to prove the existence of other, truly QM formalism
related, meaningful limits later in this treatment.

\bigskip
When generalizing the concept of QRWs in the following, we are aiming
at the following two goals:
\begin{enumerate}
\item We would like to allow for limits that, unlike
  \eqref{eq.zerolimit}, also have meaning in terms of QM-related
  descriptions of systems, beyond the limits so far obtained that have
  meaning in terms of probability theory (whereof the \emph{stationary
    limit distributions} of \eqref{eq.ahalim} were a special example,
  and the insight that QRWs are AMS added more of that kind, as pointed
  out above).
\item We would like to be able to interpret the possible existence of
  hidden states in these systems in the light of the QM formalism and
  thereby connect to earlier (well-known and largely inspiring)
  debates on the existence of hidden states within the QM formalism.
\end{enumerate}

\medskip
In this section, we make the first step towards such a unifying
clarification. We give the definition of a \emph{quantum Markov chain
  (QMC)} as a generalization of a QRW. 

\subsection{Definition}

In the following, we write

\begin{equation}
  \cV^+:=\{Q\in\cV\mid u^*Qu\ge 0\text{ for all }u\in\C^n\}
\end{equation}
for the non-negative elements of $\cV\subset\cH_n$.

\begin{definition}[Quantum Markov Chain]
  \label{def:qmc}
  Let $\cV\subset\cH_k$, $Q_0\in\cV$, $\Sigma$ be a finite set, and
  $\mu_a:\cV\to\cV, a\in\Sigma$ be $\R$-linear operators. Let
  $\mu:=\sum_a\mu_a$. We refer to the tuple
  \begin{equation}
    (\cV,(\mu_a)_{a\in\Sigma},Q_0)
  \end{equation}
  as \emph{quantum Markov chain} iff
  \begin{eqnarray}
    \label{eq:qmc:0}&&Q_0\in\cV^+\\
    \label{eq:qmc:1}&&\tr\,Q_0 = 1\\
    \label{eq:qmc:2}\text{\rm for all } Q\in\mathcal{V}:&& \tr\,\mu(Q) = \tr\,Q\\
    \label{eq:qmc:3}\text{\rm for all } a\in\Sigma:&& \mu_a(\cV^+)\subset\cV^+
  \end{eqnarray}
\end{definition}

\bigskip
\noindent {\sc Remark.}
\begin{itemize}
\item If the $\mu_a$ are \emph{completely positive},
  that is
  \begin{equation}
    (I\otimes\mu_a)(A)\ge 0\quad\text{for any nonnegative}\quad A\in
    \cW\otimes\cV
  \end{equation}
  where $\cW$ is an extra system, the $\mu_a$ are
  \emph{quantum operations} (\emph{cf.} \cite{Nielsen-Chuang}). The
  quantum operations formalism aims at modeling the dynamics of open
  quantum systems and quantum noise, borrowing from the
  interrelation between classical noise and classical Markov
  chains. Time-discrete quantum Markovian dynamics have also been
  described by trace-preserving quantum operations as \emph{quantum
    channels} (see, \emph{e.g.}, \cite{Wilde}).
\item If the $\mu_a$ reflect quantum operations, as described above,
  then the collection
  \begin{equation}
    \{\mu_a\mid a\in\Sigma\}
  \end{equation}
  is also referred to as \emph{measurement model} in the literature,
  see \cite{Nielsen-Chuang}.
\item So, when requiring the $\mu_a$ to be completely positive, the
  QMCs provide a means for extending those formalisms towards a
  clearer view on their (potential) hidden states and their temporal
  dynamics, hence their asymptotic, ergodic properties. While we could
  be happy to postulate complete positivity---which would not
  interfere with any of the following theoretical results---we refrain
  from explicitly doing so, for the sake of a clearer technical
  exposition.
\end{itemize}

\bigskip
Let $\mu_v:=\mu_{v_t}...\mu_{v_1}$ for $v=v_1...v_t$ [note the reverse
  order on the letters]. Quantum Markov chains can be seen to give
rise to stochastic processes $p$ (viewed as process functions, see
section~\ref{sec.sp}) by the rule
\begin{equation}
  p(v):=\tr\,\mu_v(Q_0)
\end{equation}
The definining properties immediately imply that
\begin{equation}
  \tr\,\mu_vQ_0\in[0,1]\quad\text{ for all }v\in\Sigma^*
\end{equation}
since [the first equation will follow from multinomial expansion]
\begin{equation}
  \begin{split}
    \sum_{v\in\Sigma^t}\tr\,\mu_vQ_0=\tr\,\mu^tQ_0\stackrel{\eqref{eq:qmc:2}}{=}\tr\,Q_0\stackrel{\eqref{eq:qmc:1}}{=}1.
  \end{split}
\end{equation}
This, by further means of (\ref{eq:qmc:0},\ref{eq:qmc:3}) shows that
\begin{equation}
  [\tr\,\mu_vQ_0]_{v\in\Sigma^t}
\end{equation}
establishes a probability distribution over $\Sigma^t$.  We recall
\eqref{eq.sp1},\eqref{eq.sp2},\eqref{eq.sp3} of section~\ref{sec.sp}
to see that $p$ corresponds to a stochastic process.

\medskip
We list some relationships of the elements of quantum Markov chains
with existing concepts of Markov chain theory and quantum mechanics
in the following subsections.

\subsection{Unitary Evolution}
\label{ssec.qmcs-and-unitaryevolution}
In quantum mechanics, evolution is described by application of unitary
matrices $U\in\C^{n\times n}$ to wave functions $\phi\in\C^n$ (that is 
$||\phi||=1$), which results in a new wave function $U\phi$ where
$||U\phi||=1$ since $U$ is unitary.  In terms of densities
$Q=\psi\psi^*$, this translates into the computation
\begin{equation}
  Q \mapsto UQU^*
\end{equation}
which establishes a linear, non-negative and trace-preserving
operation $\mu_U:\cH_n\to\cH_n$. Time-discrete, unitary evolution can
therefore be modeled in form of QMCs
\begin{equation}
  (\cV,(\mu_a)_{a\in\Sigma},Q_0)\quad\text{where}\quad |\Sigma|=1
\end{equation}
such that
\begin{equation}
  \mu=\mu_a:\cV\to\cV,Q\mapsto UQU^*
\end{equation}
with unitary $U$ describing evolution of the system.

\subsection{Measurements}
\label{ssec.qmcs-and-measurements}
A (positive operator valued) quantum measurement (= POVM, \emph{cf.}
\cite{B-NGJ,Nielsen-Chuang}) is given by a finite collection
$X=\{M_a\mid a\in \Sigma\}$ of matrices $M_a\in \C^{n\times n}$ such
that the self-adjoint matrices $X_a=M_aM_a^*$ are non-negative and sum
up to the identity:
\begin{equation}
    I = \sum_{a\in \Sigma} X_a = \sum_{a\in \Sigma}M_aM_a^*.
\end{equation}
POVMs give rise to QMCs by raising linear operators
\begin{equation}
  \mu_a:\cH_n\to\cH_n, Q\mapsto M_aQM_a^*
\end{equation}
together with a quantum density $Q_0$.  Approving the defining
principles of QMCs then is an easy exercise.

\subsection{Quantum Random Walks}
\label{ssec.qmcs-and-qrws}

\paragraph{\bf Aharonov et al.'s Model.}
QRWs $\cQ=(G,U,\psi_0)$ are seen to be QMCs, by first
introducing projection operators [we remind that
  $k=|X|\cdot|\Sigma|$ is the number of edges of the
  walk]
\begin{equation*}
  P_a:\C^k\longrightarrow\C^k,\;\psi\mapsto\sum_{(a,x),x\in X}\psi_{(a,x)}\ve_{(a,x)}
\end{equation*}
which project vectors onto the subspace spanned by basis vectors
(which are in a one-to-one correspondence with the edges) associated
with the letter $a$, and further setting
\begin{enumerate}
\item $\cV:=\cH_k$
\item $Q_0:=\psi_0\psi_0^*\in\C^{k\times k}$
\item $\mu_a:\cH_k\longrightarrow\cH_k,\;Q\mapsto (P_aU)Q(P_aU)^*$.
\end{enumerate}

\paragraph{\bf Szegedy's Model.}
According to Szegedy \cite{Szegedy04}, the unitary operator $U$ is
supposed to agree with a unitary operator $W_{P,Q}$, where the
probabilistic matrices $P,Q$ give rise to a \emph{bipartite random
  walk}. So, identifying $U$ in a QRW with some $W_{P,Q}$ in a QRW
$(U,\psi_0)$ yields a QRW in the style of Szegedy. This then further
translates into identifying also Szegedy-style QRWs with QMCs.

In Szegedy's model, just as in all more advanced QRW models, realizing
trajectories of observables is not necessarily in the focus.  Rather,
letting $U$ evolve for some time $t$ (which results in $U^t\psi_0$),
and making predictions about the expected behavior when trying to
realize an observable after time $t$ is of interest.

Such studies are, of course, equally covered by our treatment. In
particular, we have just pointed out that unitary evolution in itself
can be regarded as a QMC, see subsection
\ref{ssec.qmcs-and-unitaryevolution} above. However, when trying
to associate genuine stochastic processes with QRWs in a physically
natural way, then repeated measurements seem to be the only option.

\subsection{Hidden Markov Processes}
\label{ssec.qmcs-and-hmps}
Moreover, one can model hidden Markov processes as QMCs.
Therefore, we consider the space $\cD$ of diagonal matrices
$D\in\cH_n$. Let $M$ be the transition probability matrix of a hidden
Markov process. We see that 
\begin{equation}
  \mu(\diag(\pi)) = \diag(\pi^TM) \quad(\pi\in \R^n),
\end{equation}
establishes a non-negative, trace-preserving linear operator.
Decomposing $M$ into matrices $M_a$, as per $(M_a)_{ij}=e_{ia}\cdot
m_{ij}$ (see \eqref{eq.hmpdecomp}), we obtain non-negative linear
operators
\begin{equation}
  \mu_a:\cD\to\cD,\; \diag(\pi)\mapsto\diag(\pi^TM_a)
\end{equation}
where, obviously, $\sum_a\mu_a=\mu$. It is easy to see that
\begin{equation}
  (\cD, (\mu_a)_{a\in\Sigma},\diag(\pi))
\end{equation}
is a QMC whose associated stochastic process is that of the hidden
Markov process we started from.

\begin{Remark}
  \label{rem.hmps-vs-qrws}
  It is an immediate observation that there are HMPs that are not QRWs
  and vice versa.  This exposes both HMPs and QRWs as proper
  subclasses of QMCs.
\end{Remark}

\subsection{Hidden States}
\label{sec.qmcs-and-hiddenstates}

Representing HMPs as QMCs leads to a one-to-one correspondence of
hidden states with the eigenstates of the quantum density $\diag(\pi)$
of the QMC. In QRWs, the natural idea of hidden states is, in many
edge-based formulations (in particular in the early one raised by
Aharonov et al. \cite{Aharonov01} discussed above) that of the edges:
while one directly observes (sequences of) nodes, one does not
necessarily observe the path of edges that leads to the nodes
observed. Note that the definition of a QRW does not necessarily imply
a one-to-one correspondence of edge paths with sequences of
vertices---multiple edges connecting the same pair of nodes might be
possible.

\paragraph{\bf Remark.}
Note that in Szegedy's model \cite{Szegedy04} edges do no longer play
an explicit role. Nevertheless, pairs of nodes, and not nodes
themselves, correspond to basis vectors of the underlying Hilbert
spaces. So, mutatis mutandis, our considerations also apply for
Szegedy's model in the following. Pairs of nodes, as a concept that is
more general than edges, represent hidden states. Also, it is
immediately possible to measure hidden states, which corresponds to
projecting to the subspace spanned by just one pair of nodes.

\bigskip
Interestingly, this canonical idea of hidden states in QRWs leads to
the same analogy: when turning a QRW $(U,\psi_0)$ into a QMC
$(\cV,(\mu_v)_{v\in V},Q)$ as described above, edges (or, more
general, pairs of nodes), as canonical basis vectors of the underlying
Hilbert space turn out to be in a one-to-one correspondence with
eigenstates of the quantum density $Q=\psi_0\psi_0^*$.

\medskip
When modeling hidden states as eigenspaces, the natural question that
arises is whether one can access the hidden states through QM
formalism related operations. The immediate answer is yes. Let
$\vq_1,...,\vq_n$ be the (orthornormal) eigenstates of $Q$. Let
$P_i:\C^n\to\C^n$ be the operators that project vectors onto the
eigenspaces.  Since the $P_i$ are non-negative, and since
\begin{equation}
  I = \sum_iP_iP_i^*
\end{equation}
the $P_i$ are a POVM.\footnote{In fact, in more restrictive,
  classical QM formalism treatments, projections are the only formal
  description of quantum measurements.} Let
\begin{equation}
  T_i:\cH_n\to\cH_n;\; Q\mapsto P_iQP_i^*
\end{equation}
be the corresponding linear operators acting on the densities. Then
\begin{equation}
  p(i):=\tr\,T_iQ
\end{equation}
correspond to the probability to measure that the system described
by $Q$ is in hidden state $i$.

It is therefore possible to compute probability distributions on paths
of hidden states being taken, and, correspondingly, the most likely
path being taken (the \emph{Viterbi path}), just as is possible for
HMPs.

\section{Quantum Predictor Models}
\label{sec:qpms}

QMCs have made a first important step towards the unification of the
concepts of QRWs and HMPs. In fact, we have shown that both QRWs and
HMPs are (proper) subclasses of QMCs. For the relationships raised in
earlier work, which aimed at testing equivalence of processes in
particular \cite{Faigle11,Ito92}, there are some important questions
left to be answered.
\begin{itemize}
\item How do finitary processes relate with QMCs?
\item Can this relationship be expressed in QM formalism compatible terms?
\end{itemize}

As we will show in the following, these questions can be answered in a
satisfying way. In fact, finitary processes seem to be the natural,
unifying terminal of this treatment.

\medskip
In order to provide answers, we first recall a natural interpretation
of finitary processes: While (a finite number of) hidden states that
underlie the system might still exist, transition of hidden states is
decoupled from probability theory. This yields that one can no longer
compute most likely hidden states relative to the symbols
observed. Indeed, the hidden states only make part of the description
of the system---namely, if given a finitary process as per a
parametrization $((M_a)_{a\in\Sigma},\pi)$ where $M_a\in\R^{d\times
  d},\pi\in\R^d$, hidden states are in a one-to-one correspondence
with the canonical basis vectors of $\R^d$. $M=\sum_aM_a$ then is a
parametric (but not necessarily probabilistic!) description of how
they change. In non-HMP finitary processes, hidden states
\emph{remain (eternally) hidden} to the outside observer, who is not
in possession of their parametric description---the observer even
fails to compute reasonable estimates about them.

\medskip
In exchange, freeing hidden states from probability theory comes with
clear practical benefits:
\begin{itemize}
\item One can achieve dramatic reductions in
  terms of model complexity. See, for example, the (also
  aforementioned) ``probability clock'' \cite{Jaeger00}: a finite
  parametrization is only possible when not requiring transitions of
  hidden states to be probabilistic.
\item This idea was key to providing algorithmic solutions for the
  {\em identifiability problem}, see
  \cite{Faigle11,Ito92,Schoenhuth14}, for example.
\end{itemize}

We will therefore generalize the concept of QMCs to \emph{quantum
  predictor models (QPMs)}.  One can characterize QPMs as QMCs where
one is no longer guaranteed that performing measurements on hidden
states will work. Still, however, these hidden states are clearly
visible entities of the description of the system.

\medskip
We then show that the stochastic processes associated with QPMs are
precisely the finitary ones. This raises the following analogy:

\begin{enumerate}
\item The step from HMPs to finitary processes needs one to free
  hidden states from the laws of probability theory.
\item The step from QRWs to finitary processes needs one to free
  hidden states from being QM-measurable.
\end{enumerate}

\medskip
Beyond the demonstration of these analogies, we owe the reader a
theorem that QM formalism compatible, stationary limits exist. We will
do this in the frame of QPMs as well.

\subsection{Definition}

\begin{definition}[Quantum Predictor Model]
  Let $\cV\subset\cH_N$, $Q_0\in\cV$, $\Sigma$ be a finite set, and
  $\mu_a:\cV\to\cV, a\in\Sigma$ be $\R$-linear operators. Let
  $\mu:=\sum_a\mu_a$ and $\mu_v:=\mu_{v_l}\circ...\circ\mu_{v_1}$ for
  $v=v_1...v_l$. We refer to the tuple
  \begin{equation}
    (\cV,(\mu_a)_{a\in\Sigma},Q_0)
  \end{equation}
  as \emph{quantum predictor model (QPM)} iff
  \begin{eqnarray}
    \label{eq.qpm1}&&\tr\,Q_0 = 1\\
    \label{eq.qpm2}\text{\rm for all } Q\in\mathcal{V}:&& \tr\,\mu(Q) = \tr\,Q\\
    \label{eq.qpm3}\text{\rm for all } v\in\Sigma^*:&& \tr\,\mu_{v}Q_0\in [0,1]
  \end{eqnarray}
  In analogy to Markov chain theory, we refer to a QPM as \emph{stationary} iff
  \begin{equation}
    \mu(Q) = Q.
  \end{equation}
\end{definition}

For a better structural grasp, we also give the following definition.

\begin{definition}
  \label{d.generalized}
  Let $Q\in\cH_N$ and $\mu:\cV\to\cV$ where $\cV\subset\cH_N$ is a linear
  subspace.
  \begin{itemize}
  \item We refer to $Q$ as a \emph{generalized density} iff $\tr\,Q =
    1$.
  \item We refer to a trace-preserving linear operator $\mu:\cV\to\cV$
    as a \emph{generalized evolution operator}.
  \item We refer to $(\mu,Q)$ as \emph{generalized Markov chain} iff
    \begin{itemize}
    \item $Q$ is a generalized density and
    \item $\mu$ is a generalized evolution operator.
    \end{itemize}
  \end{itemize}
\end{definition}

Note immediately that generalized Markov chains contain ordinary
Markov chains, as per the arguments raised in
section~\ref{ssec.qmcs-and-hmps}.

We summarize the relationships of quantum predictor models with our
previous terms.

\begin{proposition}
  Let $(\cV,(\mu_a)_{a\in\Sigma},Q_0)$ be a QPM.
  \begin{enumerate}
  \item $(X_t)_{t\ge 1}$, given by
    $p(v_1...v_t) := 
    \P(\{X_1=v_1,...,X_t=v_t\} := \tr\,\mu_{v}Q_0$
    establishes a one-sided stochastic process.
  \item $(\mu, Q_0)$ is a generalized Markov chain in the sense of
    definition~\ref{d.generalized}.
  \item When the $\mu_a$ are \emph{non-negative} (that is, preserve $\cV_+$)
    and $Q_0$ is a
    \emph{quantum density}, the QPM is a QMC.
  \end{enumerate}
\end{proposition}

\Pf 1. follows from the fact that the combination of
\eqref{eq.qpm1},\eqref{eq.qpm2},\eqref{eq.qpm3} yield that $p$ is a
process function, 2. and 3. are trivial consequences of the respective
definitions.\qedd\\

\subsection{Quantum Predictor Models and Finitary Processes}
\label{ssec.qmps-are-finitary}

\begin{theorem}
  \label{t.qpm-fp}
  The class of finitary processes is equivalent to the one associated
  with quantum predictor models.
\end{theorem}

\medskip
\Pf We first show that processes associated with QPMs are
finitary. Let $p$ be the process associated with a QPM
$(\cV,(\mu_a)_{a\in\Sigma},Q_0)$ where $\dim \cV=d$. We choose a basis
$(Q_1,...,Q_{d})$ of $\mathcal{V}$ and consider the matrix
\begin{equation*}
\Pi := (\tr\,\mu_{v}Q_i)_{i\in\{1,...,d\},v\in\Sigma^*}\in\mathbb{R}^{d\times \Sigma^*},
\end{equation*}
which, because of the finite number of rows, has finite rank.
For all $v\in\Sigma^*$, we set
\begin{equation*}
\mu_{v}Q_0 =: \sum_{i=1}^{d}\alpha_{i,v}Q_i
\end{equation*}
to realize that
\begin{equation*}
p(vw) = \tr\,\mu_{w}\mu_{v}Q_0 = \tr\,\mu_{w}\sum_{i=1}^{d}\alpha_{i,v}Q_i = \sum_{i=1}^{d}\alpha_{i,v}\tr\,\mu_{w}Q_i
\end{equation*}
That is, the rows of $\mathcal{P}$ turn out to be linear combinations
of rows of $\Pi$, which implies that the rank of $\mathcal{P}$ is
finite.\\[2ex] 

For the other direction, let
$p(v_1...v_t):=\P(\{X_1=v_1,...,X_t=v_t\})$ be the process function of
a finitary process $(X_t)$. Let
$\mathcal{P}:=[p(vw)_{v,w\in\Sigma^*}]$ be the corresponding Hankel
matrix of finite rank $d$. Let $p_v:=(p(vw)_{w\in\Sigma^*})$ denote
a row of $\cP$. Let further
\begin{equation}
  \label{eq.rowspace}
  \cV_p:=\spann\{p_v\mid v\in\Sigma^*\}
\end{equation}
denote the {\em row space} of $\cP$, which is of dimension $d$.

According to the theory of finitary
processes (e.g.~\cite{Faigle11}), one can choose process functions
$p_i:\Sigma^*\to\R,i=1,...,d$ that span the
row space. That is, one can write each row
\begin{equation*}
  p_v = \sum_i\alpha_{v,i}p_i
\end{equation*}
as a linear combination of the $p_i$. 
We further observe that $\tau_a:\R^{\Sigma^*}\to\R^{\Sigma^*}$, defined
by $(\tau_ap)(w):=p(aw)$ establishes a linear operator on the space
of real-valued string functions. Since obviously
\begin{equation}
  \tau_ap_v = p_{va}
\end{equation}
$\tau_a$ preserves the row space of $\cP$. Building on this,
let $\alpha_{aij}\in\R$ be defined
through the relationship 
\begin{equation}
  \tau_ap_i = \sum_{j=1}^d\alpha_{aij}p_j
\end{equation}
We now set 
\begin{equation*}
D_i := \diag(0,...,0,\underset{i}{1},0,...,0)
\end{equation*}
and let $\cV:=\spann\{D_i,i=1,...d\}$ be the corresponding subspace of
$\cH_d$. We define the linear operators $\mu_a:\cV\to\cV$ by 
\begin{equation}
  \label{eq.ta}
  \mu_a(D_i) := \sum_{j=1}^d\alpha_{aij}D_j, 
\end{equation}
on the basis $(D_i)_{i=1,...,d}$ and further through linear extension
to all of $\mathcal{V}$.
Let further the coefficients $\alpha_{0i}$ be given through the
relationship [note that $p=\cP_{\epsilon}$ is an element of the row
  space]
\begin{equation*}
p =: \sum_{i=1}^d\alpha_{0i}p_i
\end{equation*}
and set
\begin{equation*}
Q_0:= diag(\alpha_{01},...,\alpha_{0d}).
\end{equation*}
We will show that the tuple $\mathcal{Q}_p:=(\mathcal{V},(\mu_a)_{a\in\Sigma},Q_0)$
yields a QMP which is equivalent to $p$.\\

We note first that the $\mu_a$ are linear operators by definition.
From
\begin{equation*}
\tr\,Q_0 = \sum_{i=1}^d\alpha_{0i} = \sum_{i=1}^d\alpha_{0i}\underset{=1}{\underbrace{p_i(\epsilon)}} = p(\epsilon) = 1
\end{equation*}
we obtain \eqref{eq.qpm1}. To show \eqref{eq.qpm2}, we compute for $Q\in\mathcal{V}$ and $\mu = \sum_a\mu_a$
\begin{equation*}
\begin{split}
\tr\,\mu (Q) &= \tr\,\sum_{a\in\Sigma}\mu_a(Q) = \sum_{a\in\Sigma}\tr\,\mu_a(Q)\\
&\stackrel{(*)}{=} \sum_{a\in\Sigma}\sum_{i=1}^d\sum_{j=1}^d\alpha_{aij}Q_{ii} = \sum_{i=1}^dQ_{ii}\sum_{a\in\Sigma}\sum_{j=1}^d\alpha_{aij}\\
&= \sum_{i=1}^dQ_{ii}\sum_{a\in\Sigma}\sum_{j=1}^d\alpha_{aij}p_j(\epsilon) = \sum_{i=1}^dQ_{ii}\sum_{a\in\Sigma}(\tau_a p_i)(\epsilon)\\
&= \sum_{i=1}^dQ_{ii}\sum_{a\in\Sigma}p_i(a)
= \sum_{i=1}^dQ_{ii} = tr\,Q
\end{split}
\end{equation*}
where (*) just reflects the linear extension of \eqref{eq.ta} [note
  that $Q=\sum_{i=1}^dQ_{ii}D_i$], and the last equation follows from
the fact that $p_i$ is associated with a stochastic process, which
implies $\sum_ap_i(a)=1$.\\

In the following, let $v=v_1...v_t\in\Sigma^t$ and
$\tau_{v}:=\tau_{v_t}\circ...\circ\tau_{v_1}$. We will show that
\begin{equation}
\label{eq.qpm4}
\tr\,\mu_{v}(Q_0) = \tr\,\mu_{v_t}\circ ...\circ \mu_{v_1}(Q_0) = p(v)\in [0,1]
\end{equation}
which yields \eqref{eq.qpm3} and the fact that the QPM emerging from
$\mathcal{Q}_p$ is equivalent with $p$, which completes the proof.

Therefore, for a word $v\in\Sigma$ and a vector $h\in\mathcal{V}_p$
(see \eqref{eq.rowspace}), we write
\begin{equation*}
\tau_{v}h = (\tau_{v}h)_ip_i,
\end{equation*}
that is, the $(\tau_{v}h)_i$ are the coefficients of the representation
of $\tau_{v}h$ over the basis $(p_i)$. We then show more generally that
\begin{equation}
\label{eq.qpm5}
(\mu_{v}Q_0)_{ii} = (\tau_{v}p)_i,
\end{equation}
which implies the claim because of
\begin{equation*}
\tr\,\mu_{v}(Q_0) = \sum_i(\tau_{v}p)_i = \sum_i(\tau_{v})_ip_i(\epsilon) = \tau_{v}p(\epsilon) = g(v).
\end{equation*} 

We finally show (\ref{eq.qpm5}) by induction over $t$.
For $t=1$ and $a\in\Sigma$ it holds that (note: $\alpha_{aij} = (\tau_ap_i)_j$)
\begin{equation*}
\begin{split}
(\mu_aQ_0)_{ii} &= \sum_{j=1}^d\alpha_{aji}\alpha_{0j} = \sum_{j=1}^d\alpha_{0j}(\tau_ap_j)_i\\
&= \sum_{j=1}^d(\tau_a(\alpha_{0j}p_j))_i = \tau_a(\sum_{j=1}^d\alpha_{0j}p_j)_i = (\tau_ap)_i,
\end{split}
\end{equation*}
which makes the start of the induction.
Let now $t\ge 1$ and $v=a_1...a_ta_{t+1}\in\Sigma^{t+1}$. Then it holds that
\begin{equation*}
\begin{split}
(\mu_{v}Q_0)_{ii} &= (\mu_{a_{t+1}}(\mu_{a_1...a_t}Q_0))_{ii} = \sum_{j=1}^d\alpha_{a_{t+1}ji}(\mu_{a_1...a_t})_{jj}\\
&\stackrel{(IV)}{=}  \sum_{j=1}^d\alpha_{a_{t+1}ji}(\tau_{a_1...a_t}p)_j = \sum_{j=1}^d(\tau_{a_{t+1}}(\tau_{a_1...a_t}p)_jp_j)_i\\
&= (\tau_{a_{t+1}}(\sum_{j=1}^d(\tau_{a_1...a_t}p)_jp_j))_i = (\tau_{v}p)_i.
\end{split}
\end{equation*}
\qed\\

\subsection{Asymptotic Convergence}

In the following we will point out that a special class of QPMs, which
contains the class of QMCs hence also QRWs have \emph{stationary limit
  densities}. So, we provide a theorem that ensures convenient
asymptotic ergodic properties for QRWs also in terms of the underlying
quantum concepts. This is what we were aiming at---we recall that the
attempt to compute \emph{stationary limit wave functions} (see
\eqref{eq.zerolimit}) did not lead to success.

\begin{definition}
  Let $\cQ=(\cV,(\mu_a)_{a\in\Sigma},Q)$ be a QPM and
  $\mu:=\sum_{a\in\Sigma}$ its evolution operators. We say that $\cQ$
  is \emph{bounded} if there is a $c\in\R$ such that
  \begin{equation}
    \(\mu^t(Q)|\mu^t(Q)\) = \tr(\mu^t(Q)^2) \leq c \quad\mbox{holds for all $t$.}
  \end{equation}
\end{definition}

\begin{proposition}
  \label{p.bounded}
  Let $\cQ=(\cV,(\mu_a)_{a\in\Sigma},Q)$ be a QMC. Then $\cQ$ is bounded. 
\end{proposition}

\Pf $\mu^t(Q)$ is a quantum density and therefore satisfies
$\tr(\mu^t(Q)^2)\leq 1$.\qedd

\medskip
We are able to raise the following theorem for bounded QPMs.

\begin{theorem}\label{t.Markov-limit}
  Let $\cQ=(\cV,(\mu_a)_{a\in\Sigma},Q)$ be a bounded QPM with
  evolution operator $\mu=\sum_a\mu_a$. Then the limit of averages
\begin{equation}\label{eq.Cesaro-limit}
   \tilde{Q} = \lim_{t\to\infty}\frac1t\sum_{k=1}^t \mu^t(Q)
\end{equation}
exists. Moreover, $\cQ=(\cV,(\mu_a)_{a\in\Sigma},\tilde{Q})$ is
stationary and if $Q$ is a quantum density, so is $\tilde{Q}$.
\end{theorem}

\medskip
If the limit (\ref{eq.Cesaro-limit}) exists, it is clear that
$\tr(\tilde{Q})=\tr(Q)=1$ holds and $\cQ$ is stationary. Moreover, if
$\mu$ preserves quantum densities, then each $\mu^t(Q)$, and therefore
each average, is a quantum density. So it remains to prove the
existence of $\tilde{Q}$. It is convenient to base the proof on the
following lemma.

\medskip
\begin{lemma}[\cite{Faigle07}]\label{l.FS} Let $V$
be a finite-dimensional normed vector space over $\C$ and consider
the linear operator $F:V\to V$. The following statements are
equivalent:
\begin{itemize}
\item[(a)] $\ov{v} = \lim_{t\to\infty}\frac1t\sum_{k=1}^{t-1}F^k(v)$
exists for all $v\in V$.
\item[(b)] For every $v\in V$, there exists some finite bound $c^*\in \R$ such that
$\|F^t(v)\|\leq c^*$ holds for all $t\geq 0$.
\end{itemize}
\end{lemma}

\medskip
\begin{sloppypar}
We want to apply Lemma~\ref{l.FS} to
\begin{equation}
  V := \spann\{\mu^tQ\mid t\ge 0\}\quad\text{with the norm}\quad \|C\|
=\sqrt{\tr(C^*C)}.
\end{equation}
To this end, we choose $t_0=0,t_1,...,t_m$ such that
$\{Q_j:=\mu^{t_j}(Q)\}$ is a basis for $V$.  Let $F:=\mu|_{V}$ be the
restriction of $\mu$ on $V$ (note that $V$ may be smaller than $\cV$,
while $F(V)\subset V$). That is,
\begin{equation}
  F\big(\sum_{j=0}^m r_j Q_j\big) := \sum_{j=0}^m r_j\mu(Q_j).
\end{equation}
Let $c$ be the bound on $\cQ$. We observe from the triangle
inequality:
\begin{equation}
  \|F^t\big(\sum_{i=1}^m r_i Q_i\big)\| \;\leq\; \sum_{i=1}^m
  |r_i|\cdot\|\mu^t(Q_i)\| = \sum_{i=1}^m |r_i|\cdot\|\mu^{t+t_j}(Q)\|
  \;\leq\; \sum_{i=1}^m |r_i|\sqrt{c} =:c^*\;.
\end{equation}
So $F$ satisfies condition (b) and hence also (a) of Lemma~\ref{l.FS},
which establishes the convergence of the averages in
Theorem~\ref{t.Markov-limit} with the choice $v=P$.\qedd
\end{sloppypar}

\medskip
\begin{corollary}\label{c.convergence}
  Let $\cQ=(\cV,(\mu_a)_{a\in\Sigma},Q)$ be a bounded QPM and
  $X:\cH_n\to \R$ any linear functional. Then
\begin{equation}
  \lim_{t\to\infty}\frac1t\sum_{k=1}^t X(\mu^t(P))  = X(\tilde{P}).
\end{equation}
\qed
\end{corollary}

Recalling Corollary~\ref{p.bounded} and combining it with the insight
that QRWs are QMCs, we obtain the following novel insight for quantum
random walks.

\begin{corollary}
  Quantum random walks, in their most general form, have stationary
  limit densities, hence stationary limit distributions, in the sense
  of Corollary~\ref{c.convergence}.
\end{corollary}

Note that the limit distributions one can derive via
corollary~\ref{c.convergence} substantially generalize the limit
distributions \eqref{eq.ahalim} raised in \cite{Aharonov01}.

\section{Relationship with Hidden States in Quantum Mechanics}
\label{sec.hiddenstates-in-qm}

Hidden states in quantum mechanics have played a prominent role in the
frame of debates on the EPR paradox~\cite{Einstein} and Bell's
inequalities~\cite{Bell64,Bell66}. For a consistent treatment, we will
rephrase the issue using our own terms.

\paragraph{\bf Remark.}
What follows is by no means supposed to be a re-interpretation of
physical reality, and in that sense it is not supposed to be
\emph{realistic} (in the lay sense of the word). The purpose of this
section is to point out an analogy between classical stochastic
process theory and the QM formalism.  The section will deal with the
consequences that one has to take into account when making the step
from proper random walk models towards finitary models. It is
important to understand that, in classical information theory,
finitary processes can be viewed as an attempt to ``save'' hidden
states. The problem is that one can no longer apply probability theory
when dealing with the ``saved'' hidden states. This means a certain
price for the flexibility that one gains with finitary processes over
(the more rigid) hidden Markov processes.

\medskip
This section is about the price one has to pay when trying to ``save''
hidden states (in Einstein's sense) when making the step from the
still QM formalism compatible Quantum Markov Chains towards the (no
longer QM formatlism compatible) Quantum Predictor Models. Similar to
classical theory, the gain in doing this is the added flexibility of
Quantum Predictor Models over Quantum Markov Chains when it comes to
considering asymptotic behavior of Quantum Random Walk like concepts.

\paragraph{\bf Notation.}
Let $\mathfrak{S}$ be a physical system and assume that there is a
finite set $\Omega=\{\omega_1,\ldots,\omega_N\}$ of \emph{hidden
  states} such that $\mathfrak S$ is (definitely) in one of the $N$
possible hidden states $\omega\in\Omega$ at any discrete time
$t=0,1,\ldots$.
We refer to a function
\begin{equation}
  \label{eq.infofunc}
  X:\Omega\to\Sigma
\end{equation}
as \emph{information function}.  Since $\Omega$ is finite, we may
assume $\Sigma$ to be finite as well. $\Sigma$ is supposed to consist
of values that one can observe \emph{via quantum measurements}.

\begin{Remark}
  \label{rem.deterministic}
  Here and in the following, we could assume that $X(\omega)$ is a
  probability distribution over $\Sigma$, in analogy to the
  probabilistic relationship between hidden states and emitted values
  (\emph{emission probabilities}) in hidden Markov processes. Thereby,
  we would not generalize any of our arguments. Note that HMPs can
  also be modeled as \emph{functions on finite Markov chains (FFMCs)}
  \cite{Ito92}, which model a deterministic relationship between
  hidden states and observed values in HMPs. Analogous arguments apply
  in our case (we refrain from making them explicit). For these good
  reasons, we do not assume a probabilistic relationship here.
\end{Remark}

\paragraph{\bf Hidden States.}
While one can observe values from $\Sigma$, this may not be possible
for the hidden states $\omega\in\Omega$. As usual, $\mathfrak S$ is
described (at a given time $t$) by a density $Q$.
Given $Q$, $X$ can be expressed in terms of a POVM $X=\{M_a\mid
a\in\Sigma\}$ where
\begin{equation}
  \sum_{a\in\Sigma}M_aM_a^* = I\quad\text{and}\quad\tr\,(M_aQM_a^*)\ge 0
\end{equation}
This establishes that $\tr\,(M_aQM_a^*), a\in\Sigma$ establishes a
probability distribution on $\Sigma$. We write $p_Q(a)$ for such
probabilities.

\medskip In analogy to the concept of hidden states raised for QPMs,
we associate hidden states with the eigenstates of the (initial, at
time $t=0$) density $Q$.  Let $P_{\omega}$ be the projection onto the
eigenspace of the hidden state $\omega$ (if we model temporal
dynamics, we fix those projections---they always refer to the initial
eigenstates). Let $q_{\omega}$ be the eigenvalue of $Q$ relative to
the eigenspace of $\omega$.  That is,
\begin{equation}
  q_{\omega}=\tr\,P_{\omega}QP^*_{\omega}
\end{equation}
where, in case of a non-negative density $Q$, the $q_{\omega}$ are
non-negative and sum up to one, which models that one can measure
them. This, however, is not necessarily the case for generalized
densities $Q$, which models that one \emph{cannot measure the hidden
  states}---there are no apparatuses that allow to do that.

When combining \emph{non-measurable hidden states} with
\emph{measurable information functions} $X$, we can see that the
measurement $M_a$ corresponds to a projection onto the subspace
spanned by the eigenspaces of $\omega$ where $X(\omega)=a$. That is,
the probability $p(a)=\tr\,M_aQM_a^*$ to observe $a$ on $Q$ can be
computed as
\begin{equation}
  p_Q(a) = \sum_{\omega:X(\omega)=a}q_{\omega} = \sum_{\omega:X(\omega)=a}\tr\,P_{\omega}QP^*_{\omega}.
\end{equation}

In case of real-valued information functions
$X:\Omega\to\Sigma\subset\R$, this implies that one can compute the
well-defined expectation
\begin{equation}
  E_Q(X) := \sum_{x\in \Sigma} x\cdot p_Q(x) = \sum_{x\in\Sigma}\;\;\sum_{\omega:X(\omega)=x}x\cdot q_{\omega}.
\end{equation}

\medskip
Using this setting, one can model the conflicts encountered in
prominent treatments referring to the EPR paradox, such as
\cite{Feynman,Scully-Walther-Schleich}, as attempts to jointly perform
measurements on information functions
$X_1:\Omega\to\Sigma_1,...,X_k:\Omega\to\Sigma_k$ such that certain
tuples $(a_1,...,a_k)\in\Sigma_1\times...\times\Sigma_k$ lead to
identification of hidden states whose eigenvalues are negative.

\medskip
To make this explicit, we give the following definition.

\begin{Def}
  \label{d.infofunction}
  We say that the $k$ information functions
  \begin{equation}
    (X_i:\Omega\to \Sigma_i)_{i=1,...,k}
  \end{equation}
  on the system $\mathfrak S$,
  reflected by the density $Q$, are
  \emph{jointly observable relative to Q} if the composite
  information function $X:\Omega\to\Sigma$ with
  \begin{equation}
    X(\omega) := (X_1(\omega),\ldots, X_k(\omega))\quad\mbox{and}\quad \Sigma :=\Sigma_1\times \ldots \times \Sigma_k
  \end{equation}
  is observable in $Q$.
\end{Def}

\medskip
If $(X_i:\Omega\to \Sigma_i)_{i=1,...,k}$ are jointly observable
relative to $Q$, so for each
$(a_1,...,a_k)\in\Sigma_1\times...\times\Sigma_k$ there is
$M_{(a_1,...,a_k)}$ such that
\begin{equation}
  p_Q(a_1,...,a_k):=\tr\,M_{a_1,...,a_k}QM_{a_1,...,a_k}^*
\end{equation}
is the probability to observe $(a_1,...,a_k)$. Note that in our
setting
\begin{equation}
  M_{a_1,...,a_k} = M_{a_1}\cdot...\cdot M_{a_k}
\end{equation}
since each of the measurements $M_a$ reflects a projection on a
subspace.

\begin{Remark}
This precisely is the benefit of our setting---it allows to have a
clear formal view on hidden states. Note again (see
Remark~\ref{rem.deterministic}) that the assumption of a probabilistic
relationship between hidden states and observed values, in the style
of HMPs, does not generalize our treatment.
\end{Remark}

The following statement is easy to verify.

\medskip
\begin{lemma}\label{l.joint-observations}
  Assume that the collection of $k$ information functions $X_1,\ldots,
  X_k$ is jointly observable in the Markov state $q$, then every
  subcollection $X_{i_1},\ldots, X_{i_m}$ is jointly observable in
  $q$. In particular, every individual information function $X_i$ is
  observable.  Moreover, if the $X_i$ are real-valued, also every
  product $X_iX_j$ is observable in $q$.

\qed
\end{lemma}

\medskip
Hence, if two information functions $X$ and $Y$ on the system $\mathfrak S$ are real-valued and jointly observable, their product $XY$ is statistically observable and has a well-defined expectation $E(XY)$.

\medskip
Clearly, in our setting, any collection of information functions is
jointly observable in any quantum density.

\medskip
In the following two subsections, we will put our approach into
context with earlier treatments.

\subsection{Bell's inequality}\label{sec:Bells-inequality}
The well-known inequality of Bell~\cite{Bell64,Bell66} takes the form
from the following lemma in our context as a statement on the
expectations of products of pairs of information functions.

\medskip
\begin{lemma}[Bell's inequality]\label{l.Bell} Let $X,Y,Z:\Omega\to\{-1,+1\}$ be
arbitrary information functions on the system $\mathfrak S$, described
by the density $Q$. If $X,Y$ and $Z$ are jointly observable relative
to $Q$, then the following inequality holds:
\begin{equation}\label{eq.Bell}
|E_Q(XY) -E_Q(YZ)| \;\leq\; 1 - E_Q (XZ) \;.
\end{equation}
\end{lemma}

{\small\medskip \Pf Any choice of $x,y,z\in \{-1,+1\}$ satisfies the
inequality $|xy -yz| \;\leq\; 1-xz$. Because of the joint observability
assumption, all the observation probabilities
$$
p_Q(x,y,z)= \Pr\{X=x,Y=y,Z=z\}
$$
are nonnegative real numbers that sum up to $1$. So we conclude
\begin{eqnarray*}
|E_Q(XY) -E_Q(YZ)| &=&\big|\sum_{x,y,z}
(xy-yz)p_Q(x,y,z)\big|
\;\leq\; \sum_{x,y,z}|xy-yz| p_Q(x,y,z)\\
&\leq& \sum_{x,y,z}(1-xz) p_Q(x,y,z)\;=\; 1 -E_Q(XZ)\;.
\end{eqnarray*}

\qed}

\medskip
Of course, Bell's inequality may be violated by information functions
that are pairwise but not jointly observable, because triples, while
not yet pairs of observables lead to identification of hidden
states. We raise the following example.  Consider a system $\mathfrak
S$ with a set
$\Omega=\{\omega_1,\omega_2,\omega_3,\omega_4,\omega_5\}$ of five
hidden states, for example, and three information functions
$X,Y,Z:\Omega \to \{-1,+1\}$ as in the following table:
\begin{equation}
  \begin{array}{c|cccccc}
    &\omega_1&\omega_2&\omega_3&\omega_4&\omega_5 \\  \hline
    X &-1&+1&-1&-1&-1 \\
    Y &+1&+1&-1&+1&-1 \\
    Z  &+1&+1&+1&-1&-1
  \end{array}
\end{equation}
One can check that $X,Y,Z$ are pairwise observable relative to the
generalized density
\begin{equation}
  Q=\diag(-1/3,1/3,1/3,1/3,1/3)
\end{equation}
and yield the product expectations
\begin{equation}
  E_Q(XY) = +1,\; E_Q(YZ) = -1/3,\; E_Q(XZ)= +1\;,
\end{equation}
which violate Bell's inequality (\ref{eq.Bell}).

\medskip
The explanation for this is that none of the value pairs from
$\{-1,+1\}\times\{-1,+1\}$ is in a one-to-one correspondence with
$\omega_1$, whose eigenvalue is negative, for any of the pairs
$(X,Y),(X,Z),(Y,Z)$ as composite information functions. However, it
holds that
\begin{equation}
  (X,Y,Z)^{-1}(-1,+1,+1) = \{\omega_1\}
\end{equation}
which puts $\omega_1$ in a one-to-one correspondence
with the value triple $(-1,+1,+1)$.

\medskip
\begin{Remark}\label{R.7}
  Experimental results seem to indicate that quantum systems may
  violate Bell's inequality (see, \emph{e.g.}, Aspect \emph{et
    al.}~\cite{Aspect}). This is sometimes interpreted as showing that
  quantum mechanics does not admit a theory with hidden variables. The
  generalized density picture makes it clear that a violation of
  Bell's inequality only shows that the system is studied in terms of
  measurements that are perhaps pairwise but not jointly
  observable. The existence of definite but hidden states is not
  excluded.  In fact, an experimentally observed violation of Bell's
  inequality suggests that one should not place \emph{a priori}
  nonnegativity restrictions on concepts of states into which a system
  can be prepared.
\end{Remark}

\subsection{Feynman's approach to the EPR paradox}
\label{sec:Feynman-EPR}

We raise another prominent example, originally put forward by
Feynman~\cite{Feynman}. This example served as an instance where the
assumption of hidden states leads to contradictions.  In this doing,
Feynman was one of the first to provide a mathematical model to
explain the Einstein, Rosen and Podolsky (EPR) paradox (see also
Scully {\it et al.}~\cite{Scully-Walther-Schleich}).

Feynman provides the example of a quantum
density $Q\in\C^{2\times 2}$ that reflects the preparation of a spin
$1/2$ system, for spin along the $+x$ and $+z$ axis. Accordingly, he
assumes the existence of 4 hidden states, which are in a
one-to-one-correspondence with the value tuples $(++),(+-),(-+),(--)$.
According to the preparation
(see~\cite{Feynman,Scully-Walther-Schleich} for details), relative
frequencies, and in the limit, probabilities for those tuples can be
realized by
\begin{eqnarray*}
P(++) &=& [1+\(\hat{\sigma}_z\) + \(\hat{\sigma}_x\)+\(\hat{\sigma}_y\)]/4\\
P(+-) &=& [1+\(\hat{\sigma}_z\) - \(\hat{\sigma}_x\)-\(\hat{\sigma}_y\)]/4\\
P(-+) &=& [1+\(\hat{\sigma}_z\) + \(\hat{\sigma}_x\)-\(\hat{\sigma}_y\)]/4\\
P(--) &=& [1-\(\hat{\sigma}_z\) - \(\hat{\sigma}_x\)-\(\hat{\sigma}_y\)]/4,
\end{eqnarray*}
where $\(\hat{\sigma}_x\),\(\hat{\sigma}_y\),\(\hat{\sigma}_z\)$ are
the Pauli spin operators.

Feynman realized that, depending on the quantum density $Q$, some of
these ``probabilities'' could be negative. For example, the situation
\begin{equation}
  \(\hat{\sigma}_x\)=\(\hat{\sigma}_y\)=\(\hat{\sigma}_z\)= 1/2,
\end{equation}
which by choosing an appropriate ($2\times 2$-dimensional) $Q$ is
possible, yields $P(++)=5/8, P(+-)=1/8, P(-+)=3/8, P(--)=-1/8$.

\begin{sloppypar}
These values arise in the course of measurements, which are expressed
by the Pauli operators.  As measurements are supposed to yield
statistically meaningful results---measuring value tuples relates to
sampling one of $(++),...,(--)$, so the $P(++),...,P(--)$, as the
limits of these sampling experiments, require statistical
interpretation. So $P$ not being a probability distribution leads to
probabilistic conflicts. In order to resolve the issue, Feynman
suggested to extend probability theory.
\end{sloppypar}

\bigskip We do not have to do this. The concept of generalized
densities leaves us with options Feynman did not have. In Feynman's
example, eigenspaces immediately correspond to spin
constellations. Our approach to Feynman's example, however, where
eigenspaces reflect hidden states, rather than other physical
entities, starts from the generalized density
\begin{equation}
  Q=\diag(5/8, 1/8, 3/8, -1/8).
\end{equation}
Here, the entries of $Q$, in particular $Q_{44}=-1/8$, are merely
parameters that serve to describe the state of the
system. Consequently, one does not need to interpret them further and
one does not need to extend probability theory. We further provide
\begin{equation}
  \begin{array}{c|cccccc}
    &\omega_1&\omega_2&\omega_3&\omega_4 \\  \hline
    X &+&+&-&-\\
    Z &+&-&+&-
  \end{array}
\end{equation}
as two information functions through which one has (potentially)
observational access to values for the different spins. One realizes
now that $X$ and $Z$ are not jointly measurable (observable). This
means that there is no measuring device by which one can determine
value pairs for $X$ and $Z$ simultaneously. However, by
\eqref{eq.infofunc}, it is easy to see that [again, let $q_{\omega}$
  be the eigenvalue corresponding to hidden state $\omega$]
\begin{eqnarray*}
  (p_X)_Q(+) & = & q_{\omega_1}+q_{\omega_2} = 3/4\\
  (p_X)_Q(-) & = & q_{\omega_3}+q_{\omega_4} = 1/4
\end{eqnarray*}
and, similarly, $(p_Y)_Q(+)=1,(p_Y)_Q(-)=0$, which points out that $X$
and $Z$ are measurable relative to $Q$.

\medskip The potential benefit of our approach is to formally
integrate hidden states into system preparation. In Feynman's example,
this is not possible. Hidden states can only come to life by the
attempt to determine them via measurements. Those measurements involve
to simultaneously perform two incompatible measurements [in terms of
physics: note that the Pauli operators do not commute]. The assumption
of the measurable existence of certain value tuples---the hidden
states---resulting from two incompatible measurements leads to
interpretational conflicts in Feynman's frame, but not in ours.

\section{Conclusion}
\label{sec.conclusion}

In this treatment, we have provided models that put quantum random
walks, hidden Markov processes and finitary processes into a unifying
context.  The motivation for doing so was the earlier insight that not
only hidden Markov processes, but also quantum random walks are
finitary, which yielded efficient tests for equivalence and ergodicity
also for quantum random walks. Since hidden states play a key role in
these issues, our models provide a clear, formal access to such hidden
states, now also in quantum random walks and their natural,
quantum-style generalizations. The benefits of this are twofold:
first, we have become able to re-visit hidden states in quantum
mechanics also in the light of principles that apply for finitary
processes (decoupling hidden states from probability theory), which
can allow to (dramatically) reduce model complexity. Second, this line
of research has pointed out how to obtain meaningful, quantum-style
asymptotic properties for quantum random walks, and their
generalizations. Last but not least, our treatment helps to re-visit
classical treatments on hidden states in quantum mechanics.

Future work of ours is to further explore the benefits of finitary
processes in the context of quantum information theory. Since finitary
processes both capture classical Markovian processes and quantum
computing related Markovian-style processes, further unifying insights
should be possible.

\bibliographystyle{splncs}

\end{document}